\documentclass[11pt,a4paper,twocolumn]{article}

\usepackage{amssymb,amsmath,amsthm,amsfonts}
\usepackage[english]{babel}
\usepackage[latin1]{inputenc}
\usepackage{graphicx}
\usepackage[small]{titlesec}

\newcommand{\fig}[1]{Figure~\ref{fig:#1}}

\newcommand{\captionfonts}{\footnotesize}

\makeatletter  
\long\def\@makecaption#1#2{%
  \vskip\abovecaptionskip
  \sbox\@tempboxa{{\captionfonts #1: #2}}%
  \ifdim \wd\@tempboxa >\hsize
    {\captionfonts #1: #2\par}
  \else
    \hbox to\hsize{\hfil\box\@tempboxa\hfil}%
  \fi
  \vskip\belowcaptionskip}
\makeatother   

\title{An artifact in fits to conic-based surfaces}
\author{Alfonso Pérez-Escudero, Carlos Dorronsoro, Susana Marcos}
\date{\today}

\begin{document}
\maketitle
\footnotetext[0]{Instituto de Óptica, ``Daza de Valdés", Consejo Superior de Investigaciones Científicas, Serrano 121, 28006, Madrid, Spain. \\E-mail: \{alfonso,cdorronsoro,susana\}@io.cfmac.csic.es}

\abstract
It is common in Physiological Optics to fit the corneal and the lens surfaces to conic-based surfaces (usually ellipse-based surfaces), obtaining their characteristic radius of curvature and asphericity. Here we show that the variation in radius and asphericity due to experimental noise is strongly correlated. This correlation is seen both in experimental data of the corneal topographer Pentacam and in simulations. We also show that the effect is a characteristic of the geometry of ellipses, and not restricted to any experimental device or fitting procedure.  

\section{Introduction}
\begin{figure}
	\centering		
	\includegraphics[width=.7\columnwidth]{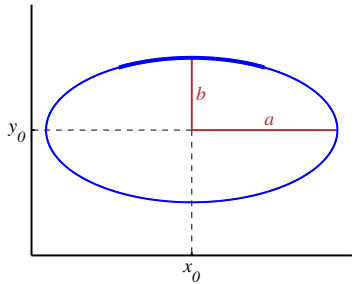}
	\caption{Ellipse centered at $(x_0, y_0)$ with semiaxes $a$, $b$ (described by Equation~\ref{ecn:ellipse}).}
	\label{fig:fig_pap_ellipse}
\end{figure}

Conic curves are the resultant of the intersection of a cone and a plane, and have simple mathematical expressions  \cite{Calossi07}. Ellipses are the most relevant conics for Physiological Optics, because both corneal and lens surfaces have usually elliptical profile. The general equation of an ellipse is
\begin{equation} \label{ecn:ellipse}
\frac{(x-x_0)^2}{a^2}+\frac{(y-y_0)^2}{b^2}=1,
\end{equation}
where $(x_0,y_0)$ is the center of the ellipse, and $a$, $b$ are the two semiaxes, \fig{fig_pap_ellipse}. Frequently (for example when we use the ellipse to describe the profile of a lens) we are not interested on the complete ellipse, but only in a part of it, usually near one apex (for example, thick region in \fig{fig_pap_ellipse}). It is convenient to describe the apical region of an ellipse in terms of two parameters, radius of curvature and asphericity \cite{Calossi07}. The radius of curvature is the radius of the circumference that best approximate the ellipse in a very small region around the apex. The asphericity is a parameter that measures how the ellipse deviates from that circumference. We can define these two parameters in terms of the two semiaxes of the ellipse, as
\begin{align}
R&=\frac{b^2}{a}\label{ecn:radius}\\
Q&=\frac{b^2}{a^2}-1 \label{ecn:asphericity},
\end{align}
where $R$ is the radius of curvature and $Q$ is the asphericity\footnote{There are several different parameters for the description of asphericity,all of them essentially equivalent \cite{Calossi07}, although see \cite{deOrtueta08}.}, and both of them are referred to the superior apex of the ellipse (thick region in \fig{fig_pap_ellipse}).

Many surfaces, such as lens surfaces, are conveniently described in terms of conic-based surfaces (most commonly, ellipse-based surfaces). One example of these kind of surfaces is the ellipsoid (\fig{fig_pap_puntos}), whose general equation is
\begin{equation} \label{ecn:ellipsoid}
\frac{(x-x_0)^2}{a^2}+\frac{(y-y_0)^2}{b^2}+\frac{(z-z_0)^2}{b^2}=1.
\end{equation}
However, this is not the only way of building a surface based on ellipses, and frequently used alternatives are, for example, biconical surfaces \cite{Schwiegerling2000,Barbero03,Cano04} and conicoids \cite{Kiely82}. Again, frequently we are interested on the region around one apex of the surface. The geometry of this region is best described in terms of the radius and asphericity of the representative meridians.

Ellipse-based surfaces parametrized in terms of radius of curvature and asphericity are widely used, for example to describe  the surface of the cornea \cite{Calossi07,Schwiegerling2000,Barbero03,Cano04,Kiely82,Llorente04,Perez-Escudero09} and the crystalline lens in the eye \cite{Dubbelman01}.

\section{Fits to conic-based surfaces}
\begin{figure}
	\centering		
	\includegraphics[width=1\columnwidth]{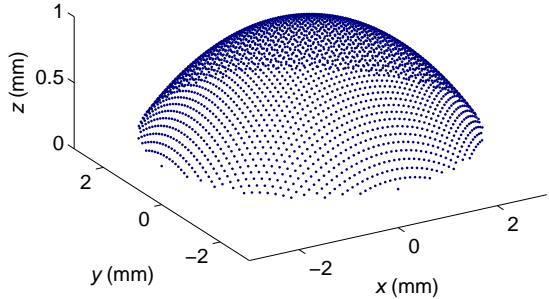}
	\caption{Computer-generated dataset. Points are on the surface of a perfect rotationally symmetric ellipsoid, with $R=6$~mm and $Q=0.3$.}
	\label{fig:fig_pap_puntos}
\end{figure}

The problem that we will discuss here refers to the fitting of the experimental measurement of a surface to an conic-based surface, obtaining its representative radius of curvature and asphericity. All the fits described in this paper were performed by minimizing the squared distances in vertical direction ($z$ axis, see \fig{fig_pap_puntos}) between the experimental points and the fitted surface. 

This paper was motivated by observations on experimental measurements taken with the corneal topographer Pentacam (Oculus GmbH, Wetzlar, Germany). Therefore, we will focus on experimental datasets in the same format as those yielded by Pentacam. These data consist of a set of points in a 3-dimensional space, that are distributed on the X-Y plane on a regular square grid of side $100\ \mu\text{m}$, \fig{fig_pap_puntos}.

\section{Correlation between fitted radius of curvature and asphericity}
\begin{figure}
	\centering		
	\includegraphics[width=1\columnwidth]{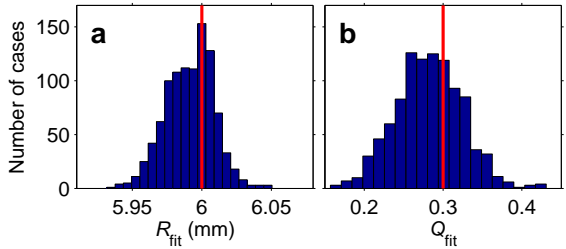}
	\caption{Results of the fits to computer-generated datasets of rotationally symmetric ellipsoids with $R=6$~mm and $Q=0.3$, with normal noise of variance 0.01~mm added in the $z$ direction. \textbf{(a)} Radius of curvature. \textbf{(b)} Asphericity. Both histograms are built from 1000 simulations.}
	\label{fig:fig_pap_histogramas}
\end{figure}

\fig{fig_pap_puntos} shows a computer-generated dataset, following a perfect rotationally symmetric ellipsoid, of radius $R_\text{real}=6$~mm and asphericity $Q_\text{real}=0.3$. If we fit these datapoints to an ellipsoid, we recover the right parameters, $R_\text{fit}=6$~mm and $Q_\text{fit}=0.3$. Now, we add some noise to the datapoints, to mimic experimental uncertainty. In particular, we add Gaussian noise with variance 0.01~mm to the $z$ component of each point. If we take 1000 realizations of the noise, and fit the datapoints, we obtain some dispersion in the recovered parameters, so that $R_\text{fit}=5.99\pm 0.02$~mm and $Q_\text{fit}=0.28 \pm 0.04$ (mean $\pm$ standard deviation), \fig{fig_pap_histogramas}. 

\begin{figure}
	\centering		
	\includegraphics[width=1\columnwidth]{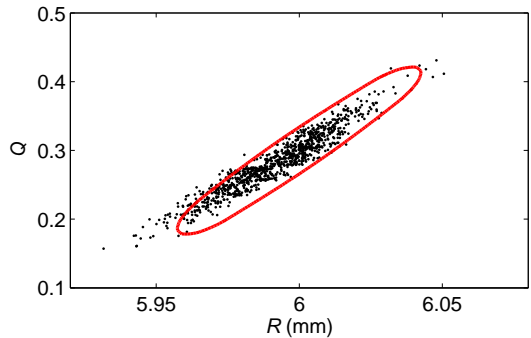}
	\caption{Parameters of fits to ellipsoids of 1000 simulations with gaussian noise added to the surface. The red contour limits the region corresponding to ellipses whose average deviation with respect to the ellipse of $R=6000$~$\mu$m and $Q=0.3$ is less than 0.5~$\mu$m.}
	\label{fig:fig_pap_correl_simul}
\end{figure}

Now, we represent the fit's asphericity versus the fit's radius of curvature, \fig{fig_pap_correl_simul}. We see a very strong and significant correlation, (\mbox{$r=0.96$}, \mbox{$p<10^{-15}$}, \mbox{$Q=-13.7+2.3R$}). We repeated this simulation with different ellipsoids and with different types of noise, getting the same correlation. Furthermore, we repeated the simulation using fits to biconics \cite{Schwiegerling2000} instead of ellipsoids, getting the same result. In order to investigate whether this correlations were intrinsically motivated by the geometrical characteristics of ellipses, we performed the following calculation: We generated ellipses\footnote{We perform this calculation with ellipses instead of ellipsoids (or any other ellipse-based surface) in order to distinguish the effects that are proper to elliptical geometry (and therefore are expected to reproduce in all ellipse-based surfaces) from those that may be particular of a certain type of ellipse-based surface.} of radii of curvature ranging between 5900 and 6200~$\mu$m, and asphericities ranging between 0.1 and 0.5, and extending 3000~$\mu$m around the apex. Then, we aligned them with respect to the ellipse with $R=6000$~$\mu$m and $Q=0.3$, so that the squared deviations were minimal. And finally we computed the average deviation between them and the ellipse with $R=6000$~$\mu$m and $Q=0.3$. As was to be expected, this average deviation is higher the more different are the parameters with respect to the reference ones. The red contour in \fig{fig_pap_correl_simul} encircles the region of the $R$-$Q$ plane for which the average deviation is lower than 0.5~$\mu$m. The good correspondence with the simulated data indicates that the correlation is due to this effect: The noise added to the surface causes the fit to find different parameters, but with higher probability for ellipses that are most similar to the correct one. As these parameters for ``similar ellipses'' are located in a region of the $R$-$Q$ plane that extends in a diagonal direction, the deviations due to the noise are correlated.

\begin{figure}
	\centering		
	\includegraphics[width=1\columnwidth]{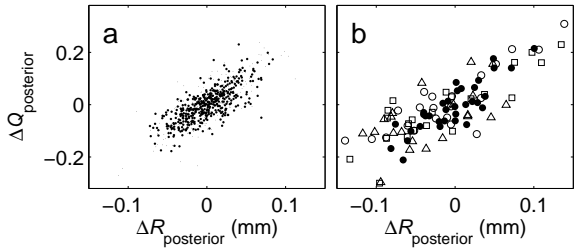}
	\caption{Results for experimental data of posterior corneal surface obtained with Pentacam. \textbf{(a)} Dispersion of measurements taken consecutively, within few minutes. \textbf{(b)} Changes between measurements taken different days. Solid dots: Non-operated controls. Empty triangles: Patients, change one day after surgery. Empty circles: Patients, change one week after surgery. Empty squares: Patients, one month after surgery.}
	\label{fig:fig_pap_correlpacientes}
\end{figure}

We have observed the effect described here in measurements of the posterior corneal surface, taken with the topographer Pentacam (Oculus) \cite{Perez-Escudero09}. \fig{fig_pap_correlpacientes}a shows the dispersion of repeated measurements taken within few minutes. Assuming that the eye does not change too much in such a short time, the observed dispersion is only attributable to experimental noise. Interestingly, the correlation also arises when these consecutive measurements are averaged, and we compute the change between measurements taken different days, \fig{fig_pap_correlpacientes}b. These measurements should be less influenced by measurement noise, due to the averaging. Thus, we might expect real changes in the cornea to be measured here, especially for eyes that correspond to patients that had undergone refractive surgery (LASIK) between the two measurements. Although many of these changes are significant (see \cite{Perez-Escudero09}), the correlation is still present. The correlation and the fact that overall dispersion is not much higher than for the case of consecutive measurements indicates that part of the effect is due to measurement noise. However, note that some points deviate clearly more than for the case of consecutive measurements. This may be indicative of some true change of the corneal geometry, but the fact that the correlation still holds indicates that the change is probably very quite small.\footnote{We carried out a reanalysis of the data presented in \cite{Perez-Escudero09}, taking into account the correlation between $R$ and $Q$. We find that, although the quantification of the changes may be biased (suggesting larger changes than those actually occuring), the distinction between significant and non-significant changes is essentially unchanged. Furthermore, as the correlation affects equally patients and control subjects (\fig{fig_pap_correlpacientes}b), the comparison between the two groups is valid. Thus, we find that the conclussins of \cite{Perez-Escudero09} are essentially unmodified by the reanalysis.}

\section{Conclusions}
Dispersion in radius of curvature and asphericity that are recovered by fitting an conic-based surface are strongly correlated. This is because the parameters that describe the most similar ellipses show the same correlation. This effect should be taken into account when we try to describe geometrical of a surface, because the true change of geometry much be much smaller (or larger) than that apparent from the change in the two parameters. Also, this effect might be used to distinguish between real changes in the geometry of the surface and changes that are apparent from the fits, but are due to measurement noise.

\end{document}